\documentclass[journal]{IEEEtran}
\usepackage{amsmath,amsfonts}
\usepackage{algorithmic}
\usepackage{algorithm}
\usepackage{array}
\usepackage[caption=false,font=normalsize,labelfont=sf,textfont=sf]{subfig}
\usepackage{lipsum}
\usepackage{booktabs} % for cmidurle
\usepackage{multirow}
\usepackage{textcomp}
\usepackage{stfloats}
\usepackage{url}
\usepackage{verbatim}
\usepackage{graphicx}
\usepackage{cite}
\DeclareMathOperator*{\argmax}{arg\,max}

\begin{document}

\title{Ask2Mask:\,Guided Data Selection for \\Masked Speech Modeling}

\author{Murali~Karthick~Baskar, Andrew~Rosenberg, Bhuvana~Ramabhadran, Yu~Zhang, Pedro~Moreno\thanks{Murali Karthick Baskar is a PhD student at Faculty of information technology, Brno university of technology, Czechia (email: baskar@vutbr.cz). Andrew Rosenberg, Bhuvana Ramabhadran, Yu Zhang and Pedro Moreno are with Google. Inc., USA}}
        % <-this % stops a space
%\thanks{This paper was produced by the IEEE Publication Technology Group. They are in Piscataway, NJ.}% <-this % stops a space
%\thanks{Manuscript received April 19, 2021; revised August 16, 2021.}}

%{Shell \MakeLowercase{\textit{et al.}
%}%: A Sample Article Using IEEEtran.cls for IEEE Journals}


\maketitle

\begin{abstract}
Masked speech modeling (MSM) methods such as wav2vec2 or w2v-BERT learn representations over speech frames which are randomly masked within an utterance. While these methods improve performance of Automatic Speech Recognition (ASR) systems, they have one major limitation.
They treat all unsupervised speech samples with equal weight, which hinders learning as not all samples have relevant information to learn meaningful representations.
In this work,  we address this limitation. We propose ask2mask (ATM), a novel approach to focus on specific samples during MSM pre-training. 
ATM employs an external ASR model or \textit{scorer} to weight unsupervised input samples in two different ways: 
1) A fine-grained data selection is performed by masking over the highly confident input frames as chosen by the scorer. This allows the model to learn meaningful representations.
2) ATM is further extended to focus at utterance-level by weighting the final MSM loss with the utterance-level confidence score. 
We conduct fine-tuning experiments on two well-benchmarked corpora: LibriSpeech (matching the pre-training data) and Commonvoice, TED-LIUM, AMI and CHiME-6 (not matching the pre-training data). The results substantiate the efficacy of ATM on significantly improving the recognition performance under mismatched conditions (up to 11.6\% relative over published results and upto 4.46\% relative over our internal baseline) while still yielding modest improvements under matched conditions.
\end{abstract}


\section{Introduction}
\IEEEPARstart{S}{elf-training} and self-supervised training techniques rely on huge amounts of unlabeled speech or text data for better generalization. The self-training techniques such as pseudo-labeling~\cite{scudder1965probability,kahn2020self} and student-teacher training~\cite{park2020improved} have shown promising improvements by incorporating the data selection process. This data selection step removes pseudo-labels with less confidence as denoted  by the teacher model before feeding the input to a student model. \cite{xu2021self} shows that self-training and self-supervised training are complementary to each other and also show that self-supervised models act as good initialization for self-training techniques. 
%why MSM is used in this work over self-training. ? claim the importance with a result from a peper.
Self-supervised training~\cite{hinton1994autoencoders} is a representation learning approach which implicitly learns patterns in the data without relying on explicit labels. Masked speech modeling (MSM) is the recent and successful self-supervised learning technique, thanks to the advent of BERT~\cite{devlin2018bert} in NLP which inspired learning speech representations from masked inputs.
MSM techniques such as wav2vec2~\cite{baevski2020wav2vec}, HuBERT~\cite{hsu2021hubert} and w2v-BERT~\cite{chung2021w2v} have shown considerable gains across various down-stream speech tasks and have become the go-to models for ASR.

Unfortunately, MSM does not have a data selection scheme to discard the irrelevant input samples and instead imposes burden on the training criterion to learn the relevance of the input samples in learning meaningful representations. 
\cite{hsu2021robust} noticed the impact of not selecting relevant data from the huge amounts of unsupervised data during pre-training by showing degradation in ASR performance when fine-tuned to a target dataset with limited data.
% we can use this as footnote
% To further brief the issue, the libri-light corpus which is commonly used for pre-training MSM models is obtained from libri-vox and contains public recorded audio books. 
%These recordings have lot of nouns, uncommon vocabulary, silence, etc. which might disrupt the learning of representations required during the fine-tuning with indomain corpus.
To mitigate this constraint,~\cite{chan2021speechstew} introduced substantially more fine-tuning data related to the target dataset but did not achieve satisfactory results. 
\cite{hsu2021robust} attempted to solve this issue by heuristically selecting the data from a closed set of unsupervised speech databases or by pooling in data relevant to target dataset along with the existing pre-training dataset.
However, this data selection approach is not done within the existing pre-training dataset and it is not completely empirically motivated.

In this study, in order to break the above limitation of the MSM techniques, we propose a simple strategy named {\em ask2mask (ATM)} to incorporate data selection within a chosen pretraining dataset.
\begin{itemize}
    \item In ATM, the masking is done over the input samples or speech frames with higher confidence as determined by the scorer. This is contrary to the random selection of frames to be masked in conventional MSM models. We hypothesize that this guided selection of frames to be masked allows the model to focus on the frames which can provide meaningful representations. The scoring model used in this work is necessarily a speech recognition model trained on small amount of data and provides frame-level confidence for each input.
sssss    \item The ATM technique is further extended to exploit the confidence values provided by the scorer by directly using them to re-weight the MSM loss. We denote this approach as ATM with loss scaling (ATM+S). It allows the model training to focus on certain {\em utterances} by down scaling the utterances with low-confidences.
\end{itemize}

Similar to our work based on masking with external guidance, there is work in NLP that also benefit by incorporating masking with knowledge. In \cite{sun2019ernie}, masking is done at phrase-level segments in BERT and has shown to learn semantic dependencies. In \cite{wang2019semantic}, phonetic knowledge is injected to mask over phonetic segments to perform spectral augmentation. Phonetically motivated masking scheme is proposed in~\cite{yue21_interspeech} to improve multiple downstream speech tasks. PEGASUS~\cite{zhang2020pegasus} model masks the input text based on their ROUGE score provides better self-supervised representations for text summarization and is more closer to the idea behind our work.

Our ATM approach is primarily motivated based on the recent work by \cite{vesely2017semi} on semi-supervised learning of conventional ASR systems which shows that performing data selection at frame-level or token-level on unsupervised data provides better performance. 
The importance of pruning out the input samples at frame-level has been studied in \cite{ferreira2021data} to improve both classification and regression tasks. Few works on unsupervised learning also highlight the importance of weighting the data based on its confidence~\cite{wessel2001confidence,ren2020not,coleman2019selection}. 
We hypothesize that ATM can leverage the effect of data selection within a particular training corpus to further enhance the recognition performance of MSM techniques. 
 
 To summarize, our contributions are listed as follows:
 \begin{itemize}
     \item \textit{Novelty}: To the extent of our knowledge, ATM is the first approach to incorporate a within-corpus data selection strategy in MSM. We also show that data selection can be simply performed inside MSM by guided selection of frames to be masked using a scorer model.
     % AR: from discussion - first approach with in semi-supervised labeling to incorporate data selection and confidence.
    \item \textit{Technical contributions}:  We provide two simple strategies to incorporate data selection into MSM pretraining by applying the confidence of the scorer: 1) choosing the data at frame-level by applying guided masking 2) soft weighting the data at utterance level by scaling the MSM loss of each utterance with its corresponding confidence score. ATM is designed to be compatible to all MSM based pre-training techniques. 
    \item \textit{Empirical study}: Analysis is done to find an optimal masking percentage for ATM and we highlight the effectiveness of ATM across varying masking percentages. The importance of masking frames with high confidence is substantiated by empirically comparing it with masking low confident frames and random frames respectively.  Experiments are performed on AMI data which is from a distinct condition compared to Libri-light corpus used for MSM based pretraining. The results confirm the importance of ATM by improving the recognition performance on evaluation sets of AMI by a significant margin.
 \end{itemize}
 
    %\item Ask-to-Mask (ATM): The masking procedure in MSM is done based on the frame confidences provided by a scoring model instead of randomly sampling from the uniform distribution. 

\section{Masked speech Models (MSM)\label{sec:msm}}
In this section, we formally define the masked speech modeling (MSM) technique and brief primary instantiations including wav2vec2 and w2v-BERT. The technique can be formulated by defining input speech sequence $\mathbf{X}=[x_{1}, x_{2},..., x_{T^{'}}]$, where $x_{t}$ is the log Mel-filterbank feature vector at time $t$. $\mathbf{X}$ is sent to the feature encoder $\Phi$ to obtain the encoded representations $\mathbf{E} = \Phi(\mathbf{X})$.  The feature encoder contains convolutional layers performing subsampling at a factor of 4 and reducing the total number of frames of an utterance from $T^{'}$ to $T$ to get $\mathbf{E} = [e_{1}, e_{2},..., e_{T}]$. $\mathbf{E}$ is then sent to two parallel modules: 1) masking component, and 2) quantizer.
%The encoded representations MSM initially performs a masking operation over the input $\mathbf{X}$ which leads to $\mathbf{X}^{rm} = [x_{1}, m_{2}, x_{3},..., x_{T}]$, where $m_{t}$ is a vector of zeros denoting the masked speech frame and $\mathbf{X}^{rm}$ denotes the randomly masked input. %The MSM training involves predicting the quantized representations $q_{t}$ of masked speech frames using either contrastive loss or cross-entropy loss objective. The quantized inputs $\mathbf{Q}$ are by default obtained using the Gumbel-softmax quantizer $\mathcal{Q}$ by feeding the masked speech input, $\mathbf{Q} = \mathcal{Q}(\mathbf{X})$. The usage of quantized targets  $Q$ is beneficial the primary difference between masked

\subsection{Masking}\label{sec:rm}
The idea behind masking  input samples and predicting them was initially proposed in BERT~\cite{devlin2018bert} and later adopted to speech~\cite{baevski2020wav2vec} with modifications to suit the characteristics of speech input.
The masking is done over sets of frames or blocks $b_{1},b_{2},...,b_{K}$ and accommodates overlap between blocks. Here $K$  is the number of masked blocks in a randomly masked encoded sequence $\tilde{\mathbf{E}}$. The importance of block masking is motivated by the improvements observed in Span-BERT by ~\cite{joshi2020spanbert} and ERNIE~\cite{sun2019ernie}.  
The block $b_{k} = [i_{k},c]$, where $i_{k}$ is the starting index of the masked block and $c$ is the corresponding right context size denoting the number of consecutive speech frames. Here $i_{k}$ are randomly sampled from a uniform distribution.
%\begin{equation}
%    i_{k} \sim \mathbf{U}(0, 1),  \forall k \in K
%\end{equation}
%Here, $\mathbf{U}$ is the uniform distribution. 
It has been empirically observed by~\cite{baevski2020wav2vec} that 49\% of the frames are masked and $c=10$ is chosen as the golden ratio to attain best representation during pre-training.
\subsection{Quantizer}\label{sec:quant}
Gumbel-softmax quantizer component $\Psi$ is used to get quantized representations which act as targets for wav2vec2 and w2v-BERT models. These quantized representations align to phonetic units as described in~\cite{baevski2020wav2vec}. Each quantized vector is of $L$ dimensions which denote the number of targets or codes used in a codebook. Each incoming input $\mathbf{E}$ is projected to $L$ dimensions within the quantizer before applying the Gumbel-softmax.
% include this into the introduction
%In this work, the experimental focus is primarily on wav2vec2 and w2v-BERT and hence described in detail.

\subsection{Context Network and MSM Loss}\label{sec:w2v2}
\textit{Wav2vec2-conformer (w2v2-cfr)}: In this model type, the unmasked sequence $\mathbf{E}$ is sent to $\Psi$ to get $\mathbf{Q} = \Psi(\mathbf{E})$, where $\mathbf{Q} = [q_{1}, q_{2}, .., q_{T}]$ as described in~\cite{baevski2020wav2vec}. 
\begin{figure}[ht]
\begin{center}
\includegraphics[scale=0.4]{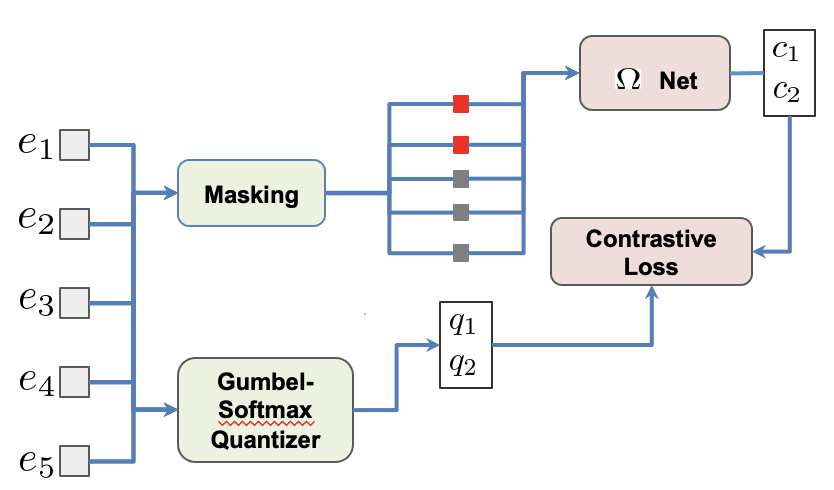}
\end{center}
\caption{Working procedure of Wav2vec2-conformer model as described in section~\ref{sec:w2v2}. The encoded representations are masked and passed to context network $\Omega$ and the resulting output $c_{j}$ is learnt to be closer to quantized output  \label{fig:wav2vec2}}
\end{figure}
\vspace{-0.9cm}
\begin{figure}[ht]
\begin{center}
\includegraphics[scale=0.4]{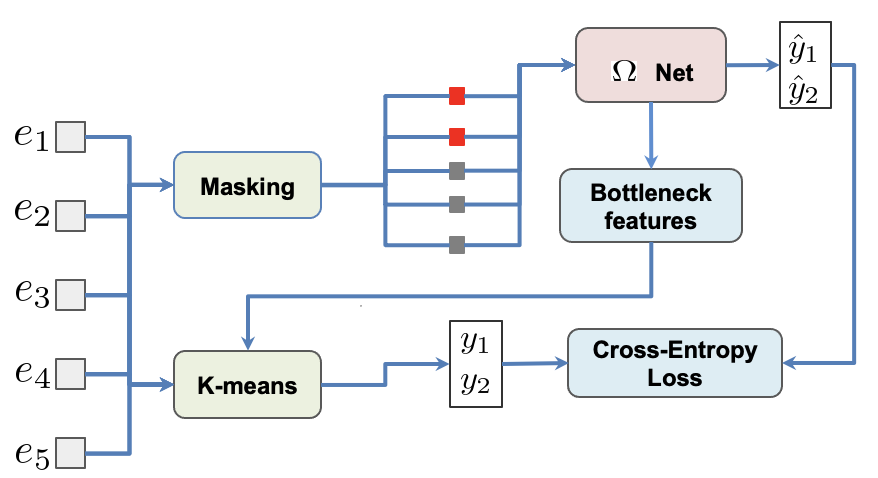}
\end{center}
\caption{Working procedure of HuBERT-conformer model as described in section~\ref{sec:w2v2}. The k-means cluster ids act as labels and they are refined using the bottleneck features extracted from the context network itself.\label{fig:hubert}}
\end{figure}
\begin{figure}[ht]
\begin{center}
\includegraphics[scale=0.4]{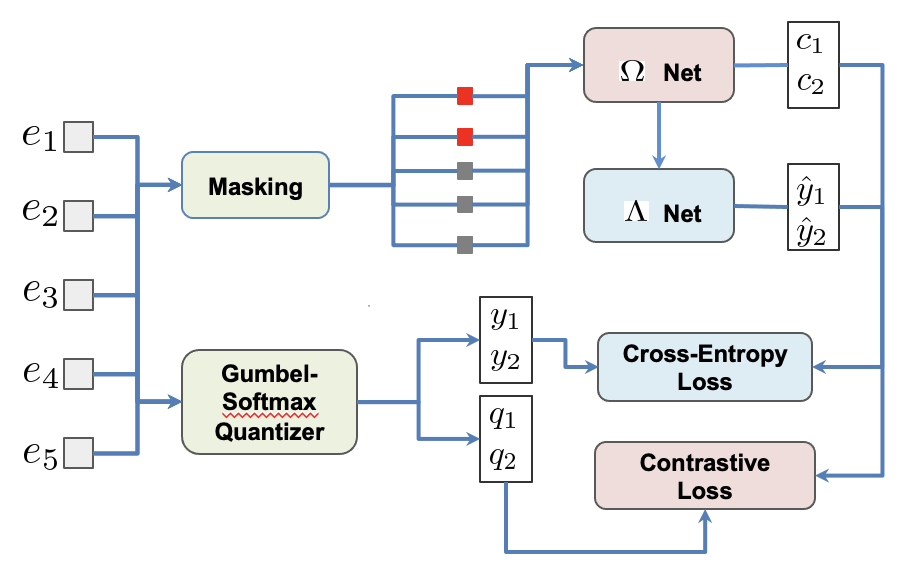}
\end{center}
\caption{Working procedure of W2V-BERT model as described in section~\ref{sec:w2vbert}. Cross-entropy loss is computed between predictions of context network $\Lambda$ and the quantized labels $y_{j}$. Contrastive loss is computed in parallel as in wav2vec2. \label{fig:w2vbert}}
\end{figure}
The masked sequence $\tilde{\mathbf{E}}$ is fed to the context network $\Omega$ which contains conformer blocks to learn contextual representations from the input. $\mathbf{C} = \Omega(\mathbf{E})$ denotes the output of the context network. 
The contrastive loss $\mathcal{L}_{ctr}(c_{j},\,q_{j})$ objective is computed between the quantized representation $q_{j}$ and context network output $c_{j} \in \mathbf{C}$ for all masked time instances $j\in J$.
Diversity loss $\mathcal{L}_{div}$ is computed as an auxiliary objective in wav2vec2 to force the model to choose diverse codes in the quantization codebook. Detailed description of $\mathcal{L}_{div}$ is in~\cite{baevski2020wav2vec}.
The final training objective is denoted as:
\begin{equation}
\mathcal{L}_{wv} = \mathcal{L}_{ctr} + 0.1 \cdot \mathcal{L}_{div},
\end{equation}
where $\mathcal{L}_{ctr} = \sum_{j=1}^{J} \mathcal{L}_{ctr}(c_{j},\,q_{j})$. 

\textit{HuBERT-conformer}: This is another variant of wav2vec2-conformer model with two major differences: 1) Targets are k-means cluster ids which are computed over a small portion of input 2) Cross-entropy loss $\mathcal{L}_{ce}(\hat{y}_{j}, \,y_{j})$ is computed between the prediction of the context network $\hat{y}_{j}$ and the k-means cluster id target $y_{j}$.

\textit{W2V-BERT}\label{sec:w2vbert}: This model marries the concept of wav2vec2 and BERT model by including an additional context network $\Lambda$ containing conformer blocks in addition to $\Omega$ as in wav2vec2. The $\Lambda$ receives the output of the $\Omega$ and strives to further learn refined contextual information to get $\mathbf{H} = \Lambda(\mathbf{C})$.
The targets of w2v-BERT $y_{j}$ is computed by taking an argmax over the codebook dimensions $L$ of quantized representations $q_{j,l}$ as:
\begin{equation}
y_{j} = \underset{l}{\argmax}\,q_{j,\,l},\, l\in L
\end{equation}
%\begin{equation}
%\hat{y}_{j,\,l} = \underset{L}{\argmax}\,softmax(c_{j,\,L}), l\in L\label{eq:2}
%\end{equation}
Finally, the cross-entropy loss $\mathcal{L}_{ce}(\hat{y}_{j},\,y_{j})$ is computed between the prediction $\hat{y}_{j}=\mathrm{softmax}(h_{j})$ and the target $y_{j}$ over the masked time instances $J$. The final training objective $\mathcal{L}_{wb} = \mathcal{L}_{ce} + \mathcal{L}_{wv}$ is a combination of cross-entropy loss and wav2vec2 loss.
%We denote these two objectives in an unified form as:
%\begin{equation}
%      \mathcal{L}_{msm}(\tilde{\mathbf{E}})=\begin{cases}
%    \mathcal{L}_{wv}, & \text{if $msm = wav2vec2$}.\\
%    \mathcal{L}_{wb}, & \text{if $msm = w2vbert$}.\label{eq:3}
%  \end{cases}
%\end{equation}
%A block diagrammatic overview of the above MSM architectures are available in appendix~\ref{app:diag}.
%Finally, there is HuBERT which is similar to wav2vec2 but replaces contrastive loss with cross-entropy loss and uses fixed targets computed using k-means. 
%\begin{wrapfigure}{R}{0.5\textwidth}
%\centering
%\begin{figure}[h]
%\begin{center}
%\includegraphics[scale=0.32]{figures/WER of pretraining using ID and OD with fine-tuning.pdf}
%\includegraphics[width=0.5\textwidth]{figures/WER on LL-60k Pretrain and speechstew finetuning.pdf}
%\framebox[4.0in]{$\;$}
%\fbox{\rule[-.5cm]{0cm}{4cm} \rule[-.5cm]{4cm}{0cm}}
%\end{center}
%\caption{\label{fig:speechstew}Performance comparison of pretraining with and without in-domain data}
%\includegraphics[width=0.25\textwidth]{frog.jpg}
%\caption{\label{fig:frog1}This is a figure caption.}
%\end{wrapfigure}

%% Please note that we have introduced automatic line number generation
%% into the style file for \LaTeXe. This is to help reviewers
%% refer to specific lines of the paper when they make their comments. Please do
%% NOT refer to these line numbers in your paper as they will be removed from the
%% style file for the final version of accepted papers.

\section{Ask2Mask (ATM)}
%\subsection{Motivation}
The primary reason to employ pre-training models is to exploit the abundantly available unsupervised data for improving ASR under limited availability of supervised data. 
While the MSM models such as wav2vec2 and w2v-BERT described in Section~\ref{sec:msm} exploit the unsupervised data, they treat each data with equal weight for computing the final objective.
% in~\Eqref{eq:3}. 
%For instance, the performance on librispeech degrades from 10.53\%WER to 14.42\%WER by pretraining only with Tedlium and Switchboard-Fischer.
Instead, we generate a score $s_{t}$ for each encoded frame $e_{t}$. This is used to select relevant data in a fine-grained manner during masking for computing the loss objective.
Here, we hypothesize that pre-training with data that closely resembles the target domain leads to better recognition performance after fine-tuning. 
% in~\Eqref{eq:3}.

%This imposes constraint on using pre-training datasets different from the target data used during evaluation. The main purpose of our proposed ask2mask approach is to carefully select the input speech frames within datasets which are more relevan higher confidences and learn representations by only training with them. 

%\begin{figure}[!h]
%\begin{center}
%\includegraphics[scale=0.6]{figures/blkb.pdf}
%\end{center}
%\caption{\label{fig:blk}Diagrammatic view of the ATM training procedure}
%\end{figure}
\subsection{Methodology}
%Figure~\ref{fig:blk} shows the working procedure of our proposed ATM approach, which employs two models: 1) scorer 2) MSM. 
For each encoded feature frame $e_{t} \in \mathbf{E}$, the scorer emits probabilities ${p(v_{t}=l\,|\,\mathbf{E}});l\in \mathbf{L}$ of the frame belonging to a particular label.
%The ATM training pipeline initially sends the input speech frames $x_{t} \in \mathbf{X}$ to the scorer model $p_{\mathbf{v}|\mathbf{x}}(v_{1:L}|x_{t})$. 
The scorer model is a CTC based frame-synchronous ASR model separately trained with a limited amount of data. Our initial intuition was to chose the scorer's training data to match the target data condition, however our empirical analysis in (cf.~Section~\ref{sec:scorer}) shows that the performance is agnostic to the scorer model's training data. 
Finally, the confidence score $s_{t}$ of the frame is defined as the maximum probability across all labels:
\begin{equation}
    s_{t} = \max_{l} p(v_{t} = l\,|\, \mathbf{E})\label{eq:5}
\end{equation}
%\begin{equation}
%    p(v_{1:L,t}) = p_{\mathbf{v}|\mathbf{x}}(x_{t})
%\end{equation}
%Here, we ask the scorer model to intake $x_{t}$ and output its probability across all labels $l=1,2,...,L$. The top-1 probability among $L$ dimensions are chosen to obtain: 
We sample $K$ masking start indices  $\{ i_{1},..,i_{k} \}$ with probabilities:
\begin{equation}
   p(i_{k} = t) = \frac{s_{t}}{\underset{v \notin \{i_{1},..,i_{k-1}\} }{\sum{s_{v}}}}\cdot \delta_{t \notin \{ i_{1}.,,i_{k-1}\}},
\end{equation}
That is, we sample beginning frames with probability proportional to the scores of each frame. The indicator function $\delta_{t\notin \{ i_{1}.,,i_{k-1}\}} $ ensures that we sample \textit{without} replacement. This is the key difference between ATM and the random masking in prior works as described in Section~\ref{sec:rm}. Prior works uniformly sample the start indices of each masking block $b_{1:K}$, while the ATM uses the probability distribution induced by the scorer. $K$ is determined by the percentage of frames to be masked.

%\begin{equation}
%    p(v_{l,t}) = \max_{\mathbf{v}}\mbox\,\,p(v_{1:L,t})
%\end{equation}
%The top-1 probability $p(v_{l,t})$ is assigned to a particular frame $x_{t}$ and this probability denotes the confidence of input frame at time $t$.
%\begin{equation}
%    p(x_{t}) = p(v_{l,t}) = \max_{\mathbf{v}}\mbox\,\,p(v_{1:L,t})
%\end{equation}
%\begin{equation}
%    i_{1}, i_{2},.., i_{K} = \max_{\mathbf{x}}K\,\,p(x_{1:T})
%\end{equation}
%Among the confidences of frames $p_{1:T}$, the top-K frames are chosen to determine the starting indices of the masking blocks $b_{1:K}$.
%The intuition to select the frames with maximum confidence stems from the thought that these frames are oriented towards the indomain distribution.
We hypothesize that frames with maximum confidence from an external scoring model will be 1) easiest to learn using an MSM training criteria and 2) most informative in for pretraining to facilitate fine-tuning.  Conversely, the lowest confidence frames, those more confusable to an external scoring model, will be the least reliably learned by MSM and least informative for pretraining.
%%
%The hypothesis to select the frames with maximum confidence is that the frames with low confidences are difficult to train with and will not allow us to learn meaningful representations.
%%

%Hence, training with these frames will allow the model to learn in-domain representations.
%The noticeable difference between random masking described in section~\ref{sec:rm} and ATM is the selection of starting indices.
%In contrast to random masking, ATM can be viewed as sampling the starting indices from the probability distribution derived by the scorer model.
%The resulting blocks of frames to be masked are then determined using $b_{k} = [i_{k},c]$.
%For instance, when $k=3$ and $c=2$ the block mask is $b_{3} = [i_{3}, i_{4}, i_{5}]$ and the resulting set of masked input frames and is denoted as $\hat{\mathbf{E}} = [e_{1}, e_{2}, m, m, m]$.
The resulting frames are sent as input to the MSM architecture and the final loss objective $\mathcal{L}$ is determined by either of the MSM objectives $\mathcal{L}_{wv}$ or $\mathcal{L}_{wb}$ described in Sections~\ref{sec:w2v2}. This modified training objective allows the model to focus on learning from gradients calculated from the frames with high confidences.

%\begin{equation}
%\label{eqn:atm}
%    \mathcal{L}_{atm} = \mathcal{L}_{msm}(\hat{\mathbf{E}})
%\end{equation}

\subsection{ATM with MSM Loss Scaling (ATM+S)}
The ATM loss %described in Equation \ref{eqn:atm}
is computed over the frames with high confidence performing a fine-grained data selection within a $u\textsuperscript{th}$ speech sequence $\mathbf{X}_{u}$. Utterances with higher average frame confidence as measured by the scorer are accorded higher value than those with more confused frames. To perform data selection at a coarser utterance level, confidence scores $s_{u}$ are computed as:
\begin{equation}
    s_{u} = \frac{1}{T}\sum_{t=1}^{T} s_{t,u}
\end{equation}
For simplicity, we denote $s = s_{u}$ and the MSM loss computed over each masked frame is scaled by $s$ to impose the importance of a particular utterance $u$. The final training objective $\mathcal{L}_{atm}^{'}$ of a particular speech sequence is denoted as:
\begin{equation}
    \mathcal{L}_{atm}^{'} = s \cdot \mathcal{L}_{atm}\label{eq:ats}
\end{equation}

%\textcolor{red}{
\subsection{Probability as confidence measure in ATM}
The ATM uses probability as a simple form of confidence measure to each frame. The analysis of confidence measures for semi-supervision in ASR has been done in~\cite{vesely2017semi} and they show that posterior probability acts as a reliable confidence measure for frame, word and sentence based data selection.
They also perform an extensive analysis on using the posteriors for hybrid ASR systems. Based on the motivation from this work we chose to use softmax probability directly as our confidence score.
A similar observation has been noted in~\cite{ferreira2021data}, where the usage of softmax probability directly as a confidence measure has been applied to select relevant data samples during training. 
We also experimented with Entropy and exponential scaling or log scaling on softmax probabilities as confidence measure, but it did not fetch advantage over simple usage of probability.
%}
\section{Experimental Setup}
All experiments including pre-training and fine-tuning are performed using 80 dimensional log Mel-filterbank features computed over the sampled 16kHz audio. Datasets (such as AMI) contains wideband audio and are downsampled to 16kHz. We  evaluate with the {test-other} (LibriSpeech partition) to show the importance of ATM on matched data conditions, while {IHM-eval} and {SDM-eval} (AMI partitions) is used to validate the model under mismatched conditions.

\subsection{Datasets used}
\textit{Pretraining (PT)}: Libri-light (LL-60k) dataset contains 60k hours of unlabeled speech and is used to pre-train all MSM models. LL-60k is the most widely used large unsupervised speech corpus for various PT techniques. Each input speech sequence is constructed by first randomly selecting 32-64 seconds segments from the original utterance. From these segments, a contiguous 32 second region is extracted from a random starting point on-the-fly during MSM PT as described in~\cite{zhang2020pushing}.

\textit{Finetuning (FT)}: Different target datasets including 1) 100 \text{\&} 960 hours of Librispeech (LS-100 \text{\&} LS-960)~\cite{panayotov2015librispeech}. 2) 100 hours of AMI and 3) speechstew (approx. 5k hours)~\cite{chan2021speechstew} are used to perform our FT experiments. Each dataset used is specific to a certain target data condition, for instance LS-960 is closely matches the LL-60k, AMI dataset is distinct from the LL-60k condition and it contains speech from two kinds of microphones ($\mathrm{i}$) Independent head microphone (IHM). ($\mathrm{ii}$) single distant microphone (SDM). SpeechStew is composed of datasets chosen from multiple conditions to create a mixed domain aggregate corpus. Processing details are described in~\cite{chan2021speechstew}. 

\textit{Evaluation}:
We hypothesize that evaluation over AMI using {IHM-eval} and {SDM-eval} reveals the effectiveness of ATM in providing informative samples for better representation learning. We also evaluate using evaluation sets from {Tedlium} and {Common voice} as their training counter parts are used in SpeechStew based FT. Finally, we also evaluate using {CHiME-6~\cite{watanabe2020chime6}} without using any FT data from {CHiME-6} training set to compare the performance of ATM on completely unseen target dataset. 

\textit{Scorer training data}:
A CTC~\cite{graves2006connectionist} based conformer model with 100M parameters is trained on LS-100 (``LS-scorer"). A similar model is also trained on AMI (``AMI-scorer"). Word-piece model (WPM) with 1024 tokens are used as labels for training the scorer models. All the results in this paper use ``LS-scorer" besides the comparison Section~\ref{sec:scorer}.

\subsection{MSM architecture}
\textit{W2v2-cfr}: This is a wav2vec2 with conformer based context network which first encodes the filterbank features  using two 2D convolutional layers with strides (2,2). 
%The quantizer contains 1024 codes quantized in the codebook which acts as targets to the final w2v2-conformer loss. 
Model has 100M/600M parameters and is denoted as ``w2v2-cfr-L/XL". HuBERT-cfr-L/XL is similar to w2v2-cfr-L/XL - it differs in using the k-means based quantizer with 1024 targets and computes the cross-entropy loss as described in~\cite{hsu2021hubert}. The ``L/XL" size models contains context network $\Omega$ 12/24 conformer layers with 8 attention heads and 1024 hidden dimensions. 

\textit{W2v-BERT}: W2v-BERT is explored using two model sizes: one with 100M parameters denoted as ``w2v-BERT-L" and containing 2 conformer layers in context net $\Omega$ and 4 conformer layers in $\Lambda$. A 600M parameter model is denoted as ``w2v-BERT-XL" contains 8 conformer layers in $\Omega$ and 24 conformer layers in $\Lambda$.  Each conformer block contains 1024 hidden dimensions with 8 attention heads, kernel size of 5 with local context of 128. The remaining architecture is identical to the configuration defined in~\cite{chung2021w2v}.
%The maximum length of input sequence is 6400 and the target sequence is 512. 
%Extra Large (XL) - 600M parameters: 
%12 layers in context net $\mathcal{C}$ and 12 layers in $\mathcal{A}$ is used for XL models.    The feature encoder filter shapes are [(3, 3, 1, 128), (3, 3, 128, 32)]
    %# 4x time reduction.
%The feature encoder filter strides = [(2, 2), (2, 2)].
%w2v2-conformer-XL: contains 24 conformer layers with 8 heads and a model dimension of 1024
%max length of input is 6400. the global batch size is  16x16x2
%We incorporate three different architectures and two different model complexity.

\subsection{PT and FT configuration}
The models L/XL are trained with a global batch size of 512/2048 on 64/256 Google TPU V3 cores for 2-4 days respectively. Adam optimizer is used with a learning rate schedule (Section 5.3 of~\cite{vaswani2017attention}) with 2e-3 as peak learning rate and 25k warmup steps. The model training configuration follows similar procedure as described in~\cite{zhang2020pushing}.

%\textcolor{red}
The FT is done by employing the context network from the PT model by adding a final layer with 1024 WPM units learnt using the RNN-T objective function~\cite{zhang2020transformer}.
The FT is done on w2v-BERT-XL, w2v2-cfr-XL and HuBERT-cfr-XL after 400k PT model updates. The w2v-BERT-L model is FT after 900k PT model updates.
%Adam optimizer with a peak learning rate of 2e-3 and 25k warmup steps is used for training all these models. The global batch size is set to 2048 on 256 TPU cores and it took 3-4 days for XL models and 2 days for the L model.
%The FT is done with a global batchsize of 1024 on 256 TPU v3 cores for 1-2 days for the L and XL models.
w2v-BERT-L is used to initially perform wide range of analysis and hyper-parameter optimization on ATM. w2v-BERT-XL is finally used to compare the results of ATM across existing works in literature.
w2v2-cfr-XL and HuBERT-cfr-XL are also used in our experiments. All these models are trained with the same configuration as in ~\cite{zhang2020pushing}.

%\subsubsection*{Baseline Models}
%\begin{table}[]
%    \centering
%    \begin{tabular}{c|c}
%         &  \\
%         & 
%    \end{tabular}
%    \caption{Caption}
%    \label{tab:baseline}
%\end{table}

\section{ATM analysis}
The empirical study on ATM is done primarily using w2v-BERT-L since this generates the best WER performance across similarly sized models (cf.~Figure~\ref{fig:consLS}). The pre-trained models are fine-tuned with either LS-100 or AMI. The resulting finetuned models are evaluated on IHM and SDM evaluation sets to understand the domain generalization aspect of ATM. Librispeech evaluation sets are used in unison to study how ATM behaves under matching domain condition. These experiments are performed with using the loss scaling (it will be discussed in Section~\ref{sec:ls}). %All our experiments which uses w2v2-conformer-XL and w2v-BERT-L/XL adopts the base configuration from existing works~\cite{chan2021speechstew,chung2021w2v,zhang2020pushing}
\begin{figure}%{R}{0.5\textwidth}
%\begin{figure}[!h]
%\begin{center}
%\includegraphics[width=0.5\textwidth]{figures/mas_percents.pdf}
\begin{centering}
\includegraphics[width=0.45\textwidth]{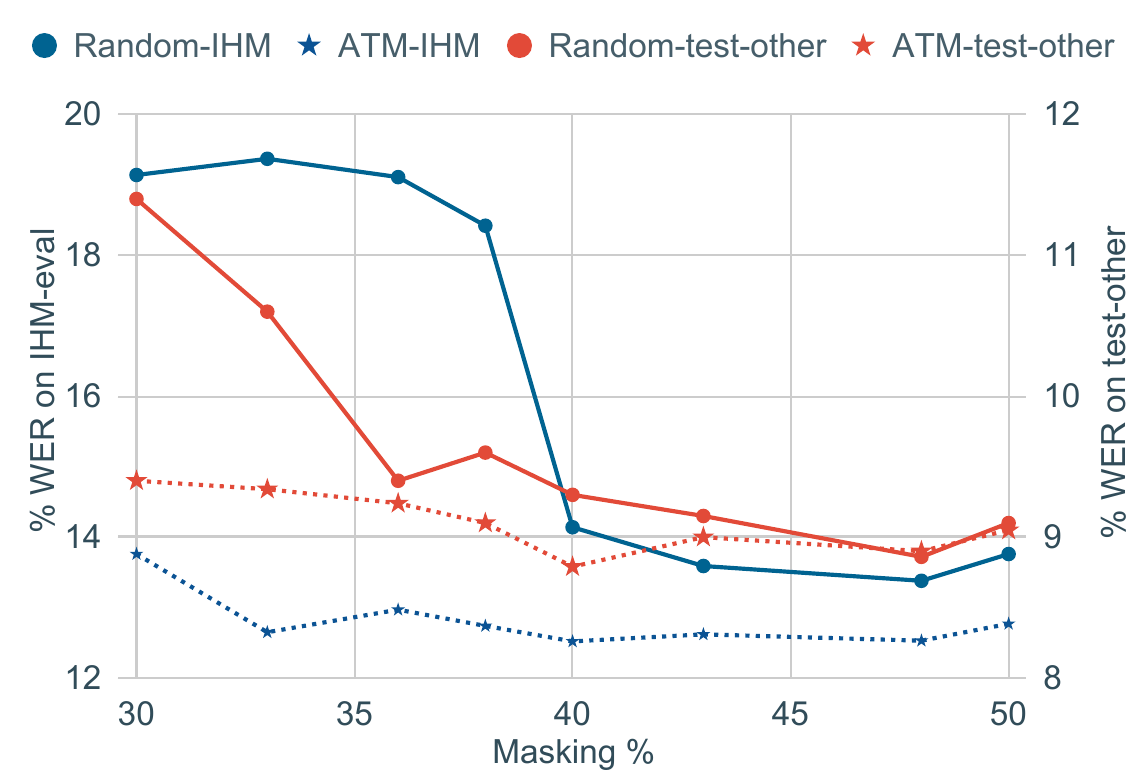}
\end{centering}
%\framebox[4.0in]{$\;$}
%\fbox{\rule[-.5cm]{0cm}{4cm} \rule[-.5cm]{4cm}{0cm}}
%\end{center}
\caption{\label{fig:percent}Recognition performance of w2v-BERT with ATM and random masking on {IHM-eval} and {test-other} sets by varying the masking percentage during pre-training. The FT is performed on LS-100 for evaluating {test-other}, while {IHM-eval} is evaluated with model FT with AMI. Random masking shows a substantial shift in performance when varying the masking from 30\% to 40\%, while the ATM remains robust to changes in masking percentage.}
%\end{figure}
\end{figure}
\subsection{Masking percentages}
The number of masked frames within an utterance plays a key role in masked input learning and in this study, we  vary the masking percentages from 30\% to 50\% to determine the best percentage for ATM approach. Previous works on wav2vec2 (\cite{baevski2020wav2vec}) showed that masking 49\%  of the frames is ideal for 30 second utterance and this has been followed subsequent works such as HuBERT and w2v-BERT.
In case of ATM, this can differ as the frames selected are of higher confidences. 
Figure~\ref{fig:percent} shows that ATM achieves its ``sweet spot'' with 40\% masking for both {IHM-eval} and {test-other} set. Interestingly ATM's performance is stable across large variations in masking rates with relatively good performance with masking rate as low as 30\%.
This is a significant difference from the uniform sampling of prior work which suffers significant drop in performance as the masking rate goes below 40\%. The result indicates that masking the right set of frames, which ATM aims to do, is able to promote more stable performance.
%A quick comparison shows that, ATM achieves better performance with very less percentage compared to random masking approach. 
For instance, ATM achieves a \%WER of 12.65 with 33\% masking and 12.52 with 40\% masking on {IHM-eval} respectively as shown in Figure~\ref{fig:percent}. 
%These results also suggest that pre-training with 40\% of high confidence speech frames selected by the scorer helps to attain better performance compared to randomly masking with 49\% of the frames.
The recognition performance on {test-other} and {IHM-eval} improves over baseline from 8.86\% to 8.79\% and 13.38\% to 12.52\% respectively by using ATM.

\begin{table}%{R}{0.5\textwidth}
\caption{Performance comparison (in \%WER) on AMI evaluation sets using w2-BERT with random masking (baseline) and with ATM using high, low and mixed confidence scores from the scorer. The FT is done with AMI.\label{tab:ls100strat}}
\begin{centering}
\resizebox{6.2cm}{!}{%
%\begin{tabular}{lccccccc}
%\hline 
%\multicolumn{2}{c}{Model} & \multicolumn{4}{c}{PT-LL, FT-LS100} & \multicolumn{2}{c}{PT-LL, FT-AMI}\tabularnewline
%\hline 
%\multicolumn{2}{c}{Info.} & dev & devother & test & test-other & IHM-eval & SDM-eval\tabularnewline
%\hline 
%\hline 
%Baseline &  & 3.78 & 8.86 & 3.85 & 9.32 & 13.38 & 31.63\tabularnewline
%%\multirow{3}{*}{ATM} & Topk & 3.71 & 8.97 & 3.89 & 8.92 & 12.52 & 27.34\tabularnewline
% & Bottomk & 3.98 & 9.59 & 4.15 & 9.76 & 14.14 & 37.77\tabularnewline
% & Mixk & 3.87 & 9.25 & 3.98 & 9.42 & 13.96 & 30.51\tabularnewline
\begin{tabular}{lccc}
\hline 
%\multicolumn{2}{c}{Model} & \multicolumn{2}{c}{PT-LL, FT-AMI}\tabularnewline
Model & Confidence  & \multicolumn{2}{c}{PT-LL, FT-AMI}\tabularnewline
%\hline 
%\multicolumn{2}{c}{Info.} & IHM-eval & SDM-eval\tabularnewline
& Level & IHM-eval & SDM-eval\tabularnewline
\hline 
\hline 
Baseline & Random & 13.38 & 31.63\tabularnewline
\hline
\multirow{3}{*}{ATM} & High & 12.52 & 27.34\tabularnewline
 & Low & 14.14 & 37.77\tabularnewline
 & Mixed & 13.96 & 30.51\tabularnewline
\hline 
\end{tabular}
}
\par \end{centering}
\end{table}
\subsection{ATM masking strategies}
The default setting of ATM is chosen based on a hypothesis that those frames that are scored with high-confidence from an external scoring model will be most useful as candidates for MSM pretraining. This hypothesis is interrogated in this section by analysing the impact of choosing the frames with low confidences or equal mix of both high and low confidence frames (Mixed). For masking low confident frames, we modify the score in ~\eqref{eq:5} as:
\begin{equation}
    s_{t} = 1 - \max_{l} p(v_{t} = l\,|\, \mathbf{E})\label{eq:6}
\end{equation}
We evaluated these three masking strategies of ATM on both IHM and SDM evaluation sets.
%\begin{figure}[!h]
%\begin{center}
%\includegraphics[scale=0.2]{figures/btm_frames.png}
%\includegraphics[scale=0.2]{figures/btm_frms.png}
%\includegraphics[scale=0.2]{figures/top_frms.png}
%\includegraphics[scale=0.2]{figures/all_frames.png}
%\framebox[4.0in]{$\;$}
%\fbox{\rule[-.5cm]{0cm}{4cm} \rule[-.5cm]{4cm}{0cm}}
%\end{center}
%\caption{Distribution of top, bottom and all frames}
%\end{figure}
%\begin{table}[!h]
%\caption{Comparison of recognition performance on LS evaluation sets using w2-BERT with random masking (baseline) and with ATM using top-k, bottom-k and mix-k strategies. The finetuning is done with LS-100\label{tab:ls100strat}}
%%\centering{}%
%\begin{tabular}{lccccc}
%\hline 
%\multicolumn{2}{c}{} & dev & devother & test & test-other\tabularnewline
%\hline 
%\hline 
%\multicolumn{2}{c}{Baseline} & 3.78 & 8.86 & 3.85 & 9.32\tabularnewline
%\hline 
%\multirow{3}{*}{ATM} & Topk & 3.71 & 8.97 & 3.89 & 8.92\tabularnewline
% & Bottomk & 3.98 & 9.59 & 4.15 & 9.76\tabularnewline
% & Mixk & 3.87 & 9.25 & 3.98 & 9.42\tabularnewline
%\hline 
%\end{tabular}
%\end{table}
Table~\ref{tab:ls100strat} shows the comparison between these sampling strategies. We observe that masking high confident frames are consistently better than masking the low confidence counterparts. In fact, ``Low" confident frames perform worse than the baseline with random masking. Finally, we observe that performance of ``Mixed" falls between that of ``High" and ``Low". The ``Mixed" strategy is similar to random masking, as both high and low confidence frames are selected.  This similarity is also reflected in comparable performance between ``Mixed'' and random masking. These results provide support for our initial hypothesis that masking frames with high confidence leads to better pre-training. 
%The performance of test-other improves from 9.32\% to 8.92\% using ATM by training with top-k frames.
%However, ATM performance on devother is slightly inferior to the baseline but remains comparable on test and dev sets. 
%This table highlights two important observations:
%1) ATM improves in the harder evaluation set such as test-other substantiating the intution behind this work
%2) Choosing top-k is consistently better compared to bottom-k and mix-k.
%\vspace{-0.3cm}
%\begin{table}[!h]
%\caption{Comparison of recognition performance on LS evaluation sets using w2-BERT with random masking (baseline) %and with ATM using top-k, bottom-k and mix-k strategies. The finetuning is done on AMI\label{tab:amistrat}}
%\begin{centering}
%\begin{tabular}{lccc}
%\hline 
% &  & IHM-eval & SDM-eval\tabularnewline
%\hline 
%\hline 
%\multicolumn{2}{c}{Baseline} & 13.38 & 31.63\tabularnewline
%\hline 
%\multirow{3}{*}{ATM} & Topk & 12.52 & 27.34\tabularnewline
% & Bottomk & 14.14 & 37.77\tabularnewline
% & Mixk & 13.96 & 30.51\tabularnewline
%\hline 
%\end{tabular}
%\par\end{centering}
%\end{table}
%We further study this effect of different masking strategies on AMI corpus as shown in table~\ref{tab:ls100strat}.
%A relative comparison between baseline and ATM shows that there is a 6.4\% relative improvement on IHM-eval using ATM and 13.6\% relative gain on SDM-eval. This trend shows that ATM performance improves with better gains as the evaluation set gets harder. 
%Bottom-k and Mix-k shows inferior performance compared to top-k on both IHM and SDM.

\subsection{How to choose the scoring model}\label{sec:scorer}
\begin{table}[h]
\caption{Cross analysis of ATM performance (in \%WER) using AMI and LS scorers. The FT is done on LS-100 to evaluate the {test} and {test-other}, while the FT is done on AMI to evaluate using IHM-eval and SDM-eval\label{tab:scorers}}
\begin{centering}
%\resizebox{5.2cm}{!}{%
\begin{tabular}{lcc}
\hline 
%\multirow{2}{*}{Evalset} & \multicolumn{2}{c}{Scorer}\tabularnewline
%\cline{2-3} \cline{3-3} 
Evalset & LS-scorer & AMI-scorer\tabularnewline
\hline
\hline
test & 3.89 & 3.93\tabularnewline
test-other & 8.92 & 9.68\tabularnewline
IHM-eval & 12.52 & 12.3\tabularnewline
SDM-eval & 27.34 & 27.00\tabularnewline
\hline 
\end{tabular}
%}
\par\end{centering}
\end{table} 
The scorer used in this work is a speech recognition model (100M parameters) trained in a supervised fashion. The scorer is chosen based on the target downstream task and in addition to this, the scorer needs to be frame-synchronous to provide confidence for every frame in a speech sequence. In this work, we use a frame-synchronous ASR system as the scorer by employing the connectionist temporal classification (CTC) objective. The CTC is preferred over the RNN-T by analysing the reliability of the frame-level predictions.
To analyse the importance of the supervised data used to train the scorer, we trained two scorer models: LS-scorer and AMI-scorer are CTC models trained with LS-100 and AMI dataset respectively.
The AMI-scorer outperforms on {SDM-eval} by improving the \%WER from 27.34 to 27.00.
Surprisingly, our results on Table~\ref{tab:scorers}, shows that the results on {IHM-eval} using an LS-scorer are comparable to the AMI-scorer.  Evaluation on {test} and {test-other} shows that LS-scorer is better than AMI-scorer on both sets. Based on these observations, we choose  the LS-scorer as the universal scoring model for all ATM based pre-trained models regardless of the target domain (eg: AMI) used in our experiments. Table~\ref{tab:scorers} shows that although matching the scorer to the target domain improves the performance, the difference is not significant.

\subsection{Consistency across different architectures}
\begin{figure}[!ht] %{R}{0.5\textwidth}
\caption{Performance comparison of different MSM architectures with and without applying ATM on IHM-eval and SDM-eval in AMI. All these models are FT using AMI. Here ``cfr" refers to conformer. \label{fig:consLS}}
\end{figure}
Figure~\ref{fig:consLS} shows that ATM consistently outperforms on both {IHM-eval} and {SDM-eval} across multiple MSM architectures including wav2vec2 and HuBERT. In the case of {IHM-eval}, ATM attains a relative improvement of 9\% over w2v2-cfr-L, 4\% relative improvement over HuBERT-cfr-L and 5\% relative gain over w2v-BERT-L baseline models respectively. W2v2-cfr-L using ATM obtained 6.2\% relative improvement over its baseline counterpart and HuBERT-cfr-L with ATM attained 7.9\% rel. improvement over HuBERT-L baseline on {SDM-eval} respectively.
%W2v-BERT-L with ATM outperformed rest of the models by attaining 13.56\% over its baseline on SDM-eval.
On the other hand w2v-BERT-L baseline is better compared to w2v2-cfr-L and HuBERT-cfr-L on both {IHM-eval} and {SDM-eval} by achieving 12.52\% and 27.34\% WER respectively.    %The w2v-BERT-XL, wav2vec2-conformer-XL and HuBERT-conformer-XL is chosen for this analysis. We chose these architectures to compare with the best possible \%WER on FT with LS-100. Table~\ref{tab:consLS} shows that ATM shows minor gains across all these architecutures on dev, devother and test. However, in test-other no gains were observed compared to the baseline model. To further analyse the behavior of the ATM, we evaluated on AMI test sets using all the three architectures and the results are in figure~\ref{fig:consLS}. .

\subsection{Analysis of ATM on Librispeech}
Experimental analysis is conducted using different model architectures to validate the effect ATM on LS-100 and are present in table~\ref{tab:consLSXL}.
The impact of increasing the model parameters from ``L" size to ``XL" size, we FT on LS-100 using MSM models with XL size and the results are in table~\ref{tab:consLSXL}. We did not find any consistency in the performance across the evaluation sets using any of the MSM architectures. Slight gains are observed on test or dev or dev-other using w2v2-cfr-XL. Once the baseline in w2v-BERT-XL gets better, ATM did not achieve gains on test-other. This scenario can be explained due to effectiveness of MSM pre-training under matched condition and can perform well without any necessary data selection approach.
\begin{table}[ht]
\caption{Performance comparison of different MSM architectures with and without applying ATM on all evaluation sets on Librispeech.\label{tab:consLSXL}}
\begin{centering}
\begin{tabular}{llcccc}
\hline 
\multirow{2}{*}{Model} & \multirow{2}{*}{Type} & \multicolumn{4}{c}{PT-LL, FT-LS100}\tabularnewline
\cmidrule(lr){3-6}
 & & dev & dev-other & test & test-other \tabularnewline
\hline 
\hline 
 w2v-BERT-L & Baseline & 3.78 & 8.86 & 3.85 & 9.32 \tabularnewline
 & ATM & 3.71 & 8.97 & 3.89 & 8.92 \tabularnewline
\cmidrule(lr){1-6} 
w2v2-cfr-XL & Baseline &  2.5 & 4.7 & 2.6 & 4.9\tabularnewline
& ATM & 2.4 & 4.6 & 2.5 & 5.0\tabularnewline
\cmidrule(lr){1-6} 
HuBERT-cfr-XL & Baseline & 2.5 & 4.7 & 2.6 & 5.0\tabularnewline
& ATM & 2.5 & 4.6 & 2.5 & 5.0\tabularnewline
\cmidrule(lr){1-6} 
w2v-BERT-XL & Baseline & 2.4 & 4.4 & 2.5 & \textbf{4.6}\tabularnewline
& ATM & \textbf{2.3} & \textbf{4.4} & \textbf{2.4} & 4.7\tabularnewline
\cmidrule(lr){1-6} 
\end{tabular}
\par\end{centering}
\end{table}

\subsection{ATM with Loss scaling (ATM+S) Analysis}\label{sec:ls}
%\begin{figure}[!h]
%\begin{center}
%\includegraphics[width=4cm, height=3.5cm]{figures/Ctrlossa.png}
%\includegraphics[width=4cm, height=3.5cm]{figures/uqa.png}
%\includegraphics[width=3.43cm, height=3cm]{figures/MLMLossa.png}
%\includegraphics[width=4cm, height=3.5cm]{figures/MLMAcca.png}
%\framebox[4.0in]{$\;$}
%\fbox{\rule[-.5cm]{0cm}{4cm} \rule[-.5cm]{4cm}{0cm}}
%\end{center}
%\caption{Validating  baseline, ATM and ATM+S using contrastive loss, number of unique codes used from the codebook and the MSM accuracy on dev-other during pre-training. A small bump is observed at around 0.1 million training iteration due to the change in learning rate and is ignored during analysis.\label{fig:ats_perfs}}
%\end{figure}
ATM training can incorporate utterance-level weighting by scaling the MSM loss obtained using w2v-BERT-L models with the utterance-level confidence score according to \eqref{eq:ats}. 
%The effect of ATM and ATM+S is analysed by plotting the validation scores on dev-other during pre-training. The first plot in figure~\ref{fig:ats_perfs} shows that the contrastive loss improves over the baseline with the aid of ATM and is further enhanced with ATM+S. The second plot shows the number of unique codes used from the quantizer codebook. 
%Analysing this plot helps us to understand if the validation loss or accuracy is improved by just using less \% of unique codes which will affect the performance at FT. 
%Among the 1024 codes, 94\%-95\% are used by both ATM and ATM+S. This is similar to the \% unique codes used by the baseline model and confirms that improvement of ATM and ATM+S is not by choosing smaller set of unique codes. The third plot shows that the MSM accuracy of ATM and ATM+S improves over baseline model. ATM+S shows that re-weighting each utterance is complementary to the ATM.
%\begin{table}[!h]
%\caption{\%WER of ATS when finetuning with LS-100}
%\centering{}%
%\begin{tabular}{ccccc}
%\hline 
% & dev & devother & test & test-other \tabularnewline
%\hline 
%\hline 
%w2v-BERT-L & 3.78 & 8.86 & 3.85 & 9.32 \tabularnewline
%ATM & 3.71 & 8.97 & 3.89 & 8.92 \tabularnewline
%+ Loss scale & 3.64 & 8.79 & 3.95 & 8.95 \tabularnewline
%\hline
%\end{tabular}
%\end{table}
\begin{table}[!ht]
\caption{\%WER of ATM+S by fine-tuning on AMI using w2v-BERT-L model\label{tab:atsami}}
\centering{}%
\begin{tabular}{clcc}
\hline 
MSM arch. & Type & IHM-eval & SDM-eval\tabularnewline
\hline 
\hline 
\multirow{2}{*}{w2v-BERT-L} & Baseline & 13.38 & 31.63\tabularnewline
 & Baseline+S & 13.14 & 28.16\tabularnewline
 & ATM & 12.52 & 27.34\tabularnewline
 & ATM+S & 13.05 & 27.19\tabularnewline
\hline
\end{tabular}
\end{table}
We evaluate the value of utterance-level loss scaling by re-weighting utterances in the context of both baseline MSM (i.e., without ATM frame selection) and ATM (ATM+S). These results are in Table~\ref{tab:atsami}. Re-weighting utterances by scaling the MSM loss with the confidence score on baseline model is denoted as ``Baseline+S" and on ATM is labeled as ``ATM+S". MSM loss scaling is effective even without ATM; baseline+S improves over baseline on both {IHM-eval} and {SDM-eval}. Moreover, ATM+S improves over ATM on {SDM-eval} by attaining 27.19\% WER while showing degradation on IHM-eval. This shows that ATM+S is effective on very hard evaluation task such as {SDM-eval} compared to {IHM-eval}. On the IHM-eval test set, the impact of MSM loss scaling is observed over the Baseline MSM without ATM. We hypothesize that ATM+S may not able to provide improvement on IHM-eval as ATM already incorporates optimally incorporates scorer information on this task. 

\begin{figure}[!ht]
\begin{center}
\includegraphics[width=4cm, height=3.5cm]{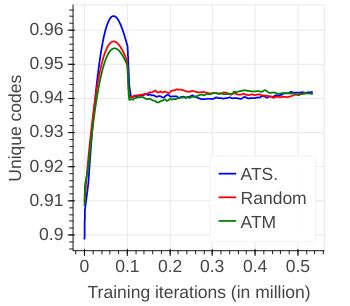}
\includegraphics[width=4cm, height=3.5cm]{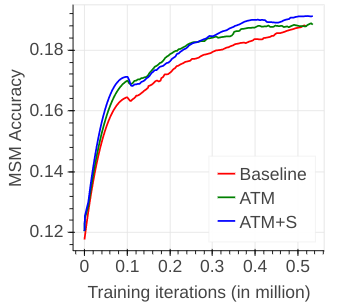}
%\framebox[4.0in]{$\;$}
%\fbox{\rule[-.5cm]{0cm}{4cm} \rule[-.5cm]{4cm}{0cm}}
\end{center}
\caption{Validating  baseline, ATM and ATM+S using contrastive loss, number of unique codes used from the codebook and the MSM accuracy on dev-other during pre-training. A small bump is observed at around 0.1 million training iteration due to the change in learning rate and is ignored during analysis.
%The bump observed in the plots is due to the sharp change in learning rate during conformer training. 
\label{fig:ats_perfs}}
\end{figure}

%You may include other additional sections here.
\subsection{ATM+S analysis on validation data during PT}
The effect of ATM and ATM+S is analysed by plotting the validation scores on dev-other during pre-training. The first plot in figure~\ref{fig:ats_perfs} shows that the contrastive loss improves over the baseline with the aid of ATM and is further enhanced with ATM+S. The second plot shows the number of unique codes used from the quantizer codebook. 
Analysing this plot helps us to understand if the validation loss or accuracy is improved by just using less \% of unique codes which will affect the performance at FT. 
Among the 1024 codes, 94\%-95\% are used by both ATM and ATM+S. This is similar to the \% unique codes used by the baseline model and confirms that improvement of ATM and ATM+S is not by choosing smaller set of unique codes. The third plot shows that the MSM accuracy of ATM and ATM+S improves over baseline model. ATM+S shows that re-weighting each utterance is complementary to the ATM.

\subsection{Comparison between frame-level and utterance-level Loss scaling }
ATM+S performs MSM loss scaling using the utterance-level confidences which performs focus on each utterance at a coarse level. We also experimented with scaling with frame-level confidence scores. Our experiments showed that scaling all utterances with frame-level confidence hurts the model performance. To solve this issue, we randomly selected utterances which participate in frame scaling. Scaling 10\% utterances resulted in better performance and the results are shown in Table~\ref{tab:framescale}.

\begin{table}[ht]
\caption{\%WER on Librispeech evaluation sets using ATM with utterance scaling and frame scaling. The frame scaling is analysed with choosing the best percentage of utterances that participate in scaling.\label{tab:framescale}}
\begin{centering}
\begin{tabular}{llcccccc}
\hline 
\multirow{2}{*}{Model} & \multirow{2}{*}{Type} & \multicolumn{4}{c}{PT-LL, FT-LS100}\tabularnewline
\cmidrule(lr){3-6}
 & & dev & dev-other & test & test-other \tabularnewline
 
\hline 
\hline 
 w2v-BERT-L & Baseline & 3.78 & 8.86 & 3.85 & 9.32\tabularnewline
% \tabularnewline
\cmidrule(lr){1-6}
\multirow{5}{*}{ATM} & None & 3.71 & 8.97 & 3.89 & \textbf{8.92} \tabularnewline
 & Utterance & \textbf{3.64} & \textbf{8.79} & 3.95 & 8.95 \tabularnewline
 & Frame-10\% & 3.73 & 8.97 & \textbf{3.84} & 9.06\tabularnewline
 & Frame-50\% & 4.01 & 9.35 & 4.14 & 9.54 \tabularnewline
 & Frame-100\% & 4.23 & 9.77 & 4.89 & 9.97 \tabularnewline
\hline 
\end{tabular}
\par\end{centering}
\end{table}

\subsection{Statistical Significance analysis on \%WER performance for multiple evaluation sets}\label{app:stasig}
The table~\ref{tab:ls960} results are on Librispeech and obtaining 0.1\% improvement in Librispeech testsets is statistically signficant. For instance, the dev-clean test set contains 54402 words and 0.1\% gains denotes a recovery of ~54 words. Also, the recent works on self-supervised training such as HuBERT shows improvement between wav2vec2-Large and HuBERT-Large only on dev-clean with 0.1\% gain in Table~\ref{tab:atsami}. 
\begin{table}[ht]
\caption{\%WER on Commonvoice using models FT with speechstew.\label{tab:commonvoice-punc}}
\begin{centering}
\begin{tabular}{lllc}
\hline 
Dataset & \%WER Imp. &  \# Total words & \# Words recovered ($\approx$) \tabularnewline
\hline 
 dev-clean & 0.1 & 54402 & 54 \tabularnewline
 test-clean & 0.1 & 52576 & 52 \tabularnewline
 AMI-IHM & 0.1 & 89635 & 89 \tabularnewline
\hline
\end{tabular}
\par\end{centering}
\end{table}

\section{Results}
In this Section, XL models are used to compare the importance of ATM on LS-960, AMI and SpeechStew. These three datasets show the effect of ATM on diverse conditions with a much larger model. Results are compared with appropriate prior work. 
%\subsection{Results on LS-960}
\begin{table}[!ht]
\caption{\%WER obtained by FT with LS-960 using w2v-BERT-XL model using baseline, ATM and ATM+S. The results show the impact of our proposed approach on matched condition since Librispeech evaluation sets are treated as closer to Libri-light PT domain.} \label{tab:ls960}
\begin{centering}
\begin{tabular}{clcccc}
\hline 
 MSM arch. & Type & dev & dev-other & test & test-other\tabularnewline
\hline 
\hline 
w2v2-cfr-XL & Baseline~\cite{chung2021w2v} & 1.7 & 3.5 & 1.7 & 3.5\tabularnewline
%\tabularnewline
\cmidrule(lr){1-6}
\multirow{2}{*}{w2v-BERT-XL} & Baseline~\cite{chung2021w2v} & 1.5 & 2.9 & 1.5 & 2.9\tabularnewline
& ATM & 1.4 & 2.8 & 1.5 & 2.9\tabularnewline
& ATM+S & 1.5 & 2.8 & 1.5 & 2.9\tabularnewline
\hline 
\end{tabular}
\par\end{centering}
\end{table}
Table~\ref{tab:ls960} shows that ATM and ATM+S improves over dev-other while on dev set there was improvement only using ATM and not ATM+S. Although the ATM and ATM+S does not show improvement on test and test-other, matches the very strong baseline.  Considering the similarity between LS-960 and PT data, ATM manages to provide gains without hurting the performance across all Librispeech evaluation sets. This validates our argument that MSM models are better without any data selection when trained under matched data condition but can benefit under mismatched conditions. 

%\subsection{Results on AMI}
\begin{table}[!ht]
\caption{\%WER obtained by FT with AMI using w2v-BERT-XL model using baseline, ATM and ATM+S. Evaluation is done on AMI test sets to highlight the effect on mismatched condition. \label{tab:amiresults}}
\begin{centering}
\begin{tabular}{clcc}
\hline 
 MSM arch. & Type & IHM-eval & SDM-eval\tabularnewline
\hline 
\hline 
%zhehuai: what's the diff between numbers inside and outside the brackets? kaldi scoring?
\multirow{2}{*}{w2v2-cfr-XL} & Baseline & 10.4 & 25.7 \tabularnewline
 & ATM & 10.0 & 24.5 \tabularnewline
& ATM+S & 9.8 & 23.9 \tabularnewline
%\multirow{2}{*}{w2v2-conformer-XL} & Baseline & 9.98 (10.4) & 21.35 (25.7) \tabularnewline
% & ATM & 9.6 (10.0) & 21.22 (24.5) \tabularnewline
%& ATM+S & 9.4 (9.8) & 21.63 (23.9) \tabularnewline
\hline 
\multirow{2}{*}{w2v-BERT-XL} & Baseline & 10.1 & 25.1\tabularnewline
 & ATM & 9.5 & 23.7 \tabularnewline
 & ATM+S & 9.5 & 23.5\tabularnewline
%\multirow{2}{*}{w2v-BERT-XL} & Baseline & 9.42  (10.1) & 21.97 (25.1)\tabularnewline
% & ATM & 9.23 (9.5) & 20.26 (23.7) \tabularnewline
% & ATM+S & 9.33 (9.5) & 19.44 (23.5)\tabularnewline
\hline 
\end{tabular}
\par\end{centering}
\end{table}
Table~\ref{tab:amiresults} presents the results of ATM on AMI by comparing it with w2v2-conformer-XL baseline and w2v-BERT-XL baselines.
We include w2v2-conformer-XL to further test the consistency of ATM on XL models when evaluated on harder tasks. 
ATM+S and ATM observes consistent gains over baseline on both IHM-eval and SDM-eval when trained with XL models.
However, ATM+S did not demonstrate improvement on IHM-eval using w2v-BERT-XL. 
%Table~\ref{tab:amiresults} also shows that ATM+S achieves 23.9\% and 23.5\% WER on SDM-eval using w2v2-conformer-XL and w2v-BERT-XL models. 
%While ATM, improves the \%WER from 25.1 to 23.7, ATM+S further pushes the performance gains. 
%Also, w2v-BERT-XL baseline is better compared to w2v2-conformer-XL baseline model which helps ATM to show additional gains. The relative improvement obtained by ATM with w2v2-conformer-XL and w2v-BERT-XL over its baseline is 4.7\% and 5.6\% on SDM-eval respectively.
%\subsubsection*{Comparison with SoTA results on speechstew}
\begin{table}[ht]
\caption{Comparison with state-of-the-art results on SpeechStew. The FT is done on SpeechStew and the results are evaluated using Kaldi scoring to match published results. 
%The Commonvoice results with $\dagger$ are obtained by scoring without punctuation as followed in~\cite{likhomanenko2020rethinking}. 
Note that the model has {\it never} seen any CHiME-6 data, and we use it as an example for \textbf{zero-shot} learning mode. \label{tab:ssresults}}
\begin{centering}
\begin{tabular}{lccccc}
\hline 
Model & Common & \multirow{2}{*}{TED} & \multicolumn{2}{c}{AMI} & \multirow{2}{*}{CHiME-6}\tabularnewline
%\cmidrule{4-5}
Type & Voice &  & IHM & SDM & \tabularnewline

\hline 
\hline 
%Speechstew &  $-$ & \multirow{2}{*}{12.1 (9.7\textsuperscript{$\dagger$})} & \multirow{2}{*}{5.3} & \multirow{2}{*}{9.0} & \multirow{2}{*}{21.7} & \multirow{2}{*}{57.2}\tabularnewline
Speechstew~\cite{chan2021speechstew}  & \multirow{1}{*}{12.1} & \multirow{1}{*}{5.3} & \multirow{1}{*}{9.0} & \multirow{1}{*}{21.7} & \multirow{1}{*}{57.2}\tabularnewline
% & & & & & &\tabularnewline
%\hline
%w2v2-conformer-XL & $-$ & \multirow{2}{*}{11.5 (9.1\textsuperscript{$\dagger$})} & \multirow{2}{*}{5.6} & \multirow{2}{*}{9.6} & \multirow{2}{*}{23.8} & \multirow{2}{*}{56.4}\tabularnewline
w2v2-cfr-XL~\cite{chan2021speechstew}  & \multirow{1}{*}{11.5} & \multirow{1}{*}{5.6} & \multirow{1}{*}{9.6} & \multirow{1}{*}{23.8} & \multirow{1}{*}{56.4}\tabularnewline
%& & & & & &\tabularnewline
%\hline
%\multirow{2}{*}{w2v-BERT-XL} & Baseline & 11.2 (9.3\textsuperscript{$\dagger$}) & 5.3 & 9.2 & 21.5 & 55.5\tabularnewline
% & ATM & 10.8 (9.2\textsuperscript{$\dagger$}) & 5.3 & 9.0 & 21.0 & 54.3\tabularnewline
% & ATM+S & 10.7 (9.1\textsuperscript{$\dagger$}) & 5.2 & 8.9 & 20.7 & 53.9\tabularnewline
\cmidrule(lr){1-6}
\multirow{1}{*}{w2v-BERT-XL} & 11.2  & 5.3 & 9.2 & 21.5 & 55.5\tabularnewline
 \,\,$+$\,ATM & 10.8  & 5.3 & 9.0 & 21.0 & 54.3\tabularnewline
 \,\,$+$\,ATM+S & 10.7 & 5.2 & 8.9 & 20.7 & 53.9\tabularnewline

\hline 
\end{tabular}
\par\end{centering}
\end{table}

%\textcolor{red}{
Table~\ref{tab:ssresults} analyses the effect of ATM and ATM+S on multiple evaluation sets such as Commonvoice, Tedlium, AMI and CHiME-6. These four sets are chosen based on the mismatch range from minimum to maximum and for instance, Commonvoice has the minimum mismatch with Libri-light data, while CHiME-6 has the maximum mismatch.
%On IHM and SDM evaluations sets, ATM+S attains 8.9\% and 20.7\% WER respectively. 
%ATM improves on SDM by attaining 21.0\% WER compared to the existing 21.7\% result on speechstew.
%}
%\textcolor{red}{
The state-of-the-art results published in~\cite{chan2021speechstew} are obtained by choosing the best Conformer model supervisedly trained with multiple datasets such as AMI, CommonVoice, Broadcast News, Librispeech, Switchboard/Fischer, TED-LIUM and Wall Street Journal. Note that the training data did not include the CHiME-6 data. The authors in~\cite{chan2021speechstew} show that simply training an ASR with lots of data leads to best results compared to the wav2vec2 finetuned model. Their best results are denoted in table~\ref{tab:ssresults} and will be used to compare with our best ATM results.
%}

%\textcolor{red}{
Our baseline w2v-BERT-XL attained better results over the published w2v2-conformer-XL and Speechstew results. In Commonvoice and CHiME-6, the baseline attained 7.4\% and 2.9\% relative improvement over Speechstew respectively. However, by including our ATM and ATM+S with w2v-BERT-XL, there was consistent improvement across all range of mismatched domains. For instance, ATM+S attains 5.76\% relative improvement on CHiME-6 over the Speechstew. This result clearly justifies that selection of reasonable input samples during pre-training reduces the necessity of having finetuning data from the same domain to improve performance. To further substantiate this, the results on AMI show a 4.6\% relative improvement on AMI-SDM over Speechstew which is of different domain compared to pre-training domain. In case of minimal mismatch domain such as Commonvoice, the ATM attained 11.6\% relative improvement over Speechstew. These observations show that ATM and ATM+S demonstrate their effectiveness to generalize to unseen and challenging speech recognition conditions.
%}

\section{Conclusion}
In this work, we introduce ask2mask (ATM) to perform data selection over unsupervised samples for MSM based pre-training to focus on relevant information and learn meaningful representations. ATM achieves 21.0\% WER on mismatched AMI SDM set with guided masking and a 20.7\% WER is obtained by including loss scaling (ATM+S). We empirically show that ATM is more robust to changes in masking percentage compared to random masking. as typically used in MSM. Our results substantiate the importance of learning from high confident frames by attaining improvements across multiple evaluation sets.
An important aspect of ATM approach is its flexibility to incorporate into any MSM pretraining techniques and ATM+S can also be easily adopted into self-supervised pre-training methods.  In our future work, we wish to apply ATM over pretraining data containing data from multiple domains~\cite{hsu2021robust,likhomanenko2020rethinking} to achieve further improvements.
We also consider two future enhancements to ATM: (1) Joint training of the scorer model with MSM model by simultaneous training on supervised and unsupervised data. (2) Perform active learning by sharing the parameters of MSM with the scorer once the MSM is well trained.

\bibliographystyle{IEEEtran}
\bibliography{references}

\begin{thebibliography}{10}
\providecommand{\url}[1]{#1}
\csname url@samestyle\endcsname
\providecommand{\newblock}{\relax}
\providecommand{\bibinfo}[2]{#2}
\providecommand{\BIBentrySTDinterwordspacing}{\spaceskip=0pt\relax}
\providecommand{\BIBentryALTinterwordstretchfactor}{4}
\providecommand{\BIBentryALTinterwordspacing}{\spaceskip=\fontdimen2\font plus
\BIBentryALTinterwordstretchfactor\fontdimen3\font minus
  \fontdimen4\font\relax}
\providecommand{\BIBforeignlanguage}[2]{{%
\expandafter\ifx\csname l@#1\endcsname\relax
\typeout{** WARNING: IEEEtran.bst: No hyphenation pattern has been}%
\typeout{** loaded for the language `#1'. Using the pattern for}%
\typeout{** the default language instead.}%
\else
\language=\csname l@#1\endcsname
\fi
#2}}
\providecommand{\BIBdecl}{\relax}
\BIBdecl

\bibitem{scudder1965probability}
H.~Scudder, ``Probability of error of some adaptive pattern-recognition
  machines,'' \emph{IEEE Transactions on Information Theory}, vol.~11, no.~3,
  pp. 363--371, 1965.

\bibitem{kahn2020self}
J.~Kahn, A.~Lee, and A.~Hannun, ``Self-training for end-to-end speech
  recognition,'' in \emph{ICASSP 2020)}.\hskip 1em plus 0.5em minus 0.4em\relax
  IEEE, 2020, pp. 7084--7088.

\bibitem{park2020improved}
D.~S. Park, Y.~Zhang, Y.~Jia, W.~Han, C.-C. Chiu, B.~Li, Y.~Wu, and Q.~V. Le,
  ``Improved noisy student training for automatic speech recognition,''
  \emph{arXiv preprint arXiv:2005.09629}, 2020.

\bibitem{xu2021self}
Q.~Xu, A.~Baevski, T.~Likhomanenko, P.~Tomasello, A.~Conneau, R.~Collobert,
  G.~Synnaeve, and M.~Auli, ``Self-training and pre-training are complementary
  for speech recognition,'' in \emph{ICASSP 2021)}.\hskip 1em plus 0.5em minus
  0.4em\relax IEEE, 2021, pp. 3030--3034.

\bibitem{hinton1994autoencoders}
G.~E. Hinton and R.~S. Zemel, ``Autoencoders, minimum description length, and
  helmholtz free energy,'' \emph{NeurIPS}, vol.~6, pp. 3--10, 1994.

\bibitem{devlin2018bert}
J.~Devlin, M.-W. Chang, K.~Lee, and K.~Toutanova, ``Bert: Pre-training of deep
  bidirectional transformers for language understanding,'' \emph{arXiv preprint
  arXiv:1810.04805}, 2018.

\bibitem{baevski2020wav2vec}
A.~Baevski, H.~Zhou, A.~Mohamed, and M.~Auli, ``wav2vec 2.0: A framework for
  self-supervised learning of speech representations,'' \emph{arXiv preprint
  arXiv:2006.11477}, 2020.

\bibitem{hsu2021hubert}
W.-N. Hsu, Y.-H.~H. Tsai, B.~Bolte, R.~Salakhutdinov, and A.~Mohamed, ``Hubert:
  How much can a bad teacher benefit asr pre-training?'' in \emph{ICASSP
  2021}.\hskip 1em plus 0.5em minus 0.4em\relax IEEE, 2021, pp. 6533--6537.

\bibitem{chung2021w2v}
Y.-A. Chung, Y.~Zhang, W.~Han, C.-C. Chiu, J.~Qin, R.~Pang, and Y.~Wu,
  ``W2v-bert: Combining contrastive learning and masked language modeling for
  self-supervised speech pre-training,'' \emph{arXiv preprint
  arXiv:2108.06209}, 2021.

\bibitem{hsu2021robust}
W.-N. Hsu, A.~Sriram, A.~Baevski, T.~Likhomanenko, Q.~Xu, V.~Pratap, J.~Kahn,
  A.~Lee, R.~Collobert, G.~Synnaeve \emph{et~al.}, ``Robust wav2vec 2.0:
  Analyzing domain shift in self-supervised pre-training,'' \emph{arXiv
  preprint arXiv:2104.01027}, 2021.

\bibitem{chan2021speechstew}
W.~Chan, D.~Park, C.~Lee, Y.~Zhang, Q.~Le, and M.~Norouzi, ``Speechstew: Simply
  mix all available speech recognition data to train one large neural
  network,'' \emph{arXiv preprint arXiv:2104.02133}, 2021.

\bibitem{sun2019ernie}
Y.~Sun, S.~Wang, Y.~Li, S.~Feng, X.~Chen, H.~Zhang, X.~Tian, D.~Zhu, H.~Tian,
  and H.~Wu, ``Ernie: Enhanced representation through knowledge integration,''
  \emph{arXiv preprint arXiv:1904.09223}, 2019.

\bibitem{wang2019semantic}
C.~Wang, Y.~Wu, Y.~Du, J.~Li, S.~Liu, L.~Lu, S.~Ren, G.~Ye, S.~Zhao, and
  M.~Zhou, ``Semantic mask for transformer based end-to-end speech
  recognition,'' \emph{arXiv preprint arXiv:1912.03010}, 2019.

\bibitem{yue21_interspeech}
X.~Yue and H.~Li, ``{Phonetically Motivated Self-Supervised Speech
  Representation Learning},'' in \emph{Proc. Interspeech 2021}, 2021, pp.
  746--750.

\bibitem{zhang2020pegasus}
J.~Zhang, Y.~Zhao, M.~Saleh, and P.~Liu, ``Pegasus: Pre-training with extracted
  gap-sentences for abstractive summarization,'' in \emph{ICML}.\hskip 1em plus
  0.5em minus 0.4em\relax PMLR, 2020, pp. 11\,328--11\,339.

\bibitem{vesely2017semi}
K.~Vesel{\'y}, L.~Burget, and J.~Cernock{\'y}, ``Semi-supervised dnn training
  with word selection for asr.'' in \emph{Interspeech}, 2017, pp. 3687--3691.

\bibitem{ferreira2021data}
J.~Ferreira, M.~Mendonca, and P.~S. Diniz, ``Data selection in neural
  networks,'' \emph{IEEE Open Journal of Signal Processing}, 2021.

\bibitem{wessel2001confidence}
F.~Wessel, R.~Schluter, K.~Macherey, and H.~Ney, ``Confidence measures for
  large vocabulary continuous speech recognition,'' \emph{IEEE Transactions on
  speech and audio processing}, vol.~9, no.~3, pp. 288--298, 2001.

\bibitem{ren2020not}
Z.~Ren, R.~Yeh, and A.~Schwing, ``Not all unlabeled data are equal: Learning to
  weight data in semi-supervised learning,'' \emph{NeurIPS}, vol.~33, 2020.

\bibitem{coleman2019selection}
C.~Coleman, C.~Yeh, S.~Mussmann, B.~Mirzasoleiman, P.~Bailis, P.~Liang,
  J.~Leskovec, and M.~Zaharia, ``Selection via proxy: Efficient data selection
  for deep learning,'' \emph{arXiv preprint arXiv:1906.11829}, 2019.

\bibitem{joshi2020spanbert}
M.~Joshi, D.~Chen, Y.~Liu, D.~S. Weld, L.~Zettlemoyer, and O.~Levy, ``Spanbert:
  Improving pre-training by representing and predicting spans,''
  \emph{Transactions of the Association for Computational Linguistics}, vol.~8,
  pp. 64--77, 2020.

\bibitem{zhang2020pushing}
Y.~Zhang, J.~Qin, D.~S. Park, W.~Han, C.-C. Chiu, R.~Pang, Q.~V. Le, and Y.~Wu,
  ``Pushing the limits of semi-supervised learning for automatic speech
  recognition,'' \emph{arXiv preprint arXiv:2010.10504}, 2020.

\bibitem{panayotov2015librispeech}
V.~Panayotov, G.~Chen, D.~Povey, and S.~Khudanpur, ``Librispeech: an asr corpus
  based on public domain audio books,'' in \emph{ICASSP 2015}.\hskip 1em plus
  0.5em minus 0.4em\relax IEEE, 2015, pp. 5206--5210.

\bibitem{watanabe2020chime6}
S.~W. et. al., ``Chime-6 challenge:tackling multispeaker speech recognition for
  unsegmented recordings,'' 2020.

\bibitem{graves2006connectionist}
A.~Graves, S.~Fern{\'a}ndez, F.~Gomez, and J.~Schmidhuber, ``Connectionist
  temporal classification: labelling unsegmented sequence data with recurrent
  neural networks,'' in \emph{ICML}, 2006, pp. 369--376.

\bibitem{vaswani2017attention}
A.~Vaswani, N.~Shazeer, N.~Parmar, J.~Uszkoreit, L.~Jones, A.~N. Gomez,
  {\L}.~Kaiser, and I.~Polosukhin, ``Attention is all you need,'' in
  \emph{NeurIPS}, 2017, pp. 5998--6008.

\bibitem{zhang2020transformer}
Q.~Zhang, H.~Lu, H.~Sak, A.~Tripathi, E.~McDermott, S.~Koo, and S.~Kumar,
  ``Transformer transducer: A streamable speech recognition model with
  transformer encoders and rnn-t loss,'' in \emph{ICASSP 2020}.\hskip 1em plus
  0.5em minus 0.4em\relax IEEE, 2020, pp. 7829--7833.

\bibitem{likhomanenko2020rethinking}
T.~Likhomanenko, Q.~Xu, V.~Pratap, P.~Tomasello, J.~Kahn, G.~Avidov,
  R.~Collobert, and G.~Synnaeve, ``Rethinking evaluation in asr: Are our models
  robust enough?'' \emph{arXiv preprint arXiv:2010.11745}, 2020.

\end{thebibliography}

%\bibliography{iclr2022_conference}
%\bibliographystyle{iclr2022_conference}
\section*{Acknowledgments}
%\subsubsection*{Acknowledgments}
We would like to thank Lukáš Burget, Jan Honza Černocký, Bolaji Yusuf and Zhehuai Chen for helping to review the paper and provide valuable suggestions.

\end{document}